\documentclass[11pt]{article}

\usepackage[final]{acl}

\usepackage{times}
\usepackage{latexsym}
\usepackage{float} % add this in your preamble
\usepackage{subcaption}

\usepackage[T1]{fontenc}
\usepackage[utf8]{inputenc}

\usepackage{microtype}

\usepackage{inconsolata}

\usepackage{graphicx}

\usepackage{amsmath, amssymb}
\usepackage{graphicx}
\usepackage{multirow}
\usepackage{natbib}
\usepackage{booktabs}

\title{Embedding the Teacher: Distilling vLLM Preferences for Scalable Image Retrieval}

\author{
\textbf{Eric He$^\natural$}\thanks{\; Research based on work during internship at Apta.} \quad
\textbf{Akash Gupta$^\S$} \quad
\textbf{Adian Liusie$^\S$} \quad
\textbf{Vatsal Raina$^\S$} \quad
\\
\textbf{Piotr Molenda$^\S$} \quad
\textbf{Shirom Chabra$^\S$} \quad
\textbf{Vyas Raina$^\S$} \\
$^\natural$University of Cambridge \quad $^\S$Apta\\
\texttt{eh733@cam.ac.uk, vyas@apta.chat}\\
}

\begin{document}
\maketitle

\begin{abstract}

Text--image retrieval is necessary for applications such as product recommendation. Embedding-based approaches like CLIP enable efficient large-scale retrieval via vector similarity search, but they are primarily trained on literal caption-like text--image pairs and often fail to capture abstract or persona-driven attributes common in product recommendation applications (e.g., ``a gift for a mother who loves gardening''). In contrast, state-of-the-art vision--language models (vLLMs) can align text with images in a flexible manner, but their limited context window prevents them from directly handling retrieval over large catalogs. We propose a framework that distills the preference rankings of a powerful vLLM into an embedding-based system, transferring its nuanced alignment abilities while maintaining the inference-time scalability of an embedding-based approach. Experiments on persona-driven product recommendation tasks demonstrate that our method significantly outperforms existing embedding-based baselines, providing an efficient solution for personalized text--image retrieval.
\end{abstract}

% \begin{abstract}

% State-of-the-art vision--language models (vLLMs) perform well at aligning text with small sets of candidate images, but they cannot scale to large image corpora due to context window limitations. Embedding-based retrieval methods such as CLIP offer efficiency at scale, but they are trained primarily for literal caption--image alignment and struggle when queries involve abstract or persona-driven attributes, as in product recommendation. We propose a method to distill vLLM preference rankings into an embedding-based system, enabling efficient vector search while preserving the nuanced alignment capabilities of the teacher vLLM. Experiments on large-scale catalog selection tasks demonstrate that our approach outperforms existing embedding-based baselines, providing a practical solution for persona-driven multi-modal recommendation.
% \end{abstract}

\section{Introduction}
    Text--image retrieval is an important task for many applications such as product recommendation systems and content platforms~\cite{youtube, recommender}.
    A common approach for such text–image retrieval tasks is to use embedding-based retrieval systems such as CLIP~\cite{CLIP}, ALIGN~\cite{jia2021scalingvisualvisionlanguagerepresentation}, or SigLIP~\cite{zhai2023sigmoidlosslanguageimage}. 
    These models project text and images into a shared embedding space, enabling efficient retrieval via vector similarity search. While effective for particular domains, these methods are trained on bespoke large-scale datasets (typically image captioning style), which may limit their generalization to other domains~\cite{visual-rag, yuan2021florencenewfoundationmodel,luo2021clip4clipempiricalstudyclip}.
    For example, if a user asks: ``My mum loves gardening, which Tiffany \& Co. bracelet would she enjoy most?'', a standard embedding-based system may struggle to align with abstract attributes (e.g., ``loves gardening''), and instead focus primarily on literal properties typical of image captioning datasets, such as object categories and colors.

\begin{figure}[t]
    \centering
    \includegraphics[width=1\linewidth]{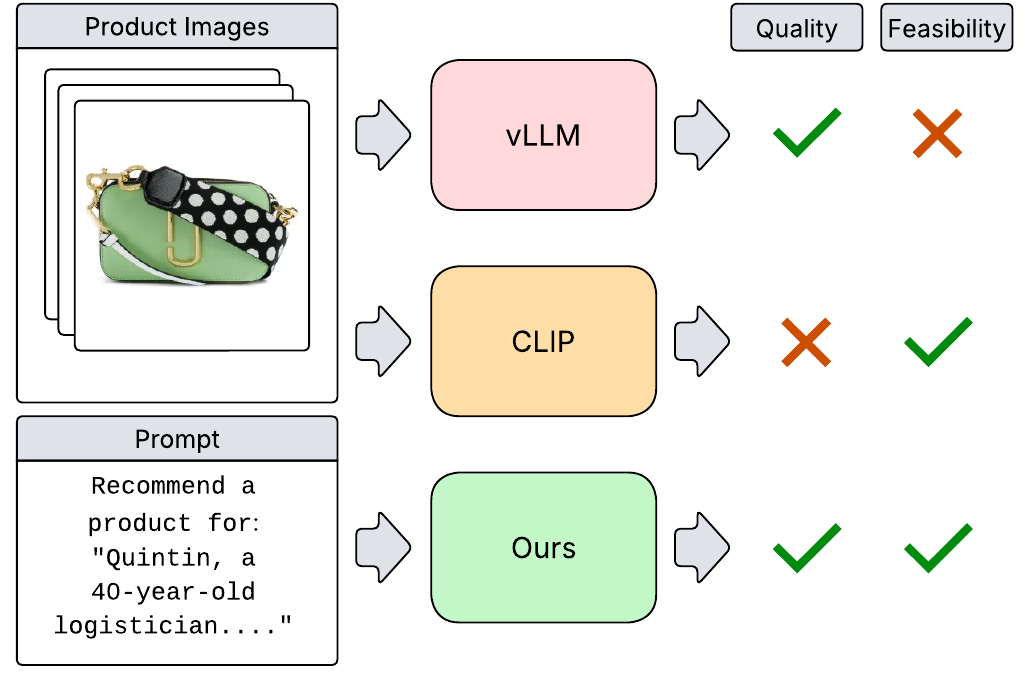}
    \caption{Comparison of methods for persona-based product image retrieval. vLLMs capture preferences well but cannot retrieve from a large image catalog (context limit); Embedding-based CLIP enables retrieval but is not well aligned to product recommendation. Our approach distills vLLM preferences into an embedding-based approach to achieve both.}
    \label{fig:teaser}
\end{figure}

    Vision--language models (vLLMs) have shown strong performance in a wide range of multi-modal tasks, often even in a zero-shot fashion~\cite{Gemini, Qwen, Llama}. 
    These models, though, have  a limitation on the size of their input context-window size, and so can not be directly used for large-scale text--image retrieval, where there can often be thousands or millions of candidate images.
    vLLMs, however, are a viable and effective approach when the number of candidate images is small. Therefore this paper considers how we can leverage the flexibility and of vLLMs to perform text--image retrieval with any specific text prompt style of interest, but in a manner that is computationally efficient and does not require any manual data labeling. In this work we consider product recommendation tasks, where the query is a rich natural-language description of a persona and the candidate images are drawn from extensive product catalogs~\cite{MileBench}.
    
    We propose a practical approach to address the challenge of aligning text--image retrieval to the property of interest (persona-based recommendation). Our approach (Figure~\ref{fig:teaser}) is to efficiently \emph{distill} the image preference rankings of a powerful vLLM into a lightweight embedding-based model. By supervising the embedding model with pairwise ranking signals derived from the vLLM, we enable large-scale, efficient recommendation through simple similarity search while retaining the nuanced, persona-driven alignment capabilities of the teacher vLLM. This approach addresses both scalability, by enabling fast inference over large catalogs, and adaptability, as it requires no manually labeled dataset of persona-to-product matches. Our implementation is available at \footnote{\url{https://github.com/ericyh/Preference_Aligned_Distillation}}.

\section{Method}

    \subsection{Problem Setup}

    We consider the task of selecting the most suitable image from a large catalog given a natural language description. The description, \(x\) may be abstract or persona-driven (e.g., \textit{``a mother who loves gardening''}), while the catalog $\mathcal{U} = \{u_1, u_2, \ldots, u_N\}$ may contain thousands of product images (e.g., Tiffany bracelets). The goal is to identify a single image $\hat{u} \in \mathcal{U}$ that best matches the description.

    A powerful vision--language model (vLLM) can be prompted to compare pairs of images and determine the more suitable one for a given description. In principle, the best image $\hat{u}$ can be obtained through a tournament-style elimination procedure requiring $N-1$ pairwise calls. However, for large $N$ this approach is computationally infeasible, motivating the need for a more efficient solution. We therefore propose to train an embedding-based model that approximates the vLLM’s rankings, enabling efficient retrieval over large catalogs.

    \subsection{Efficient Inference via Embeddings}
    
\begin{figure}[t]
        \centering
        \includegraphics[width=0.9\linewidth]{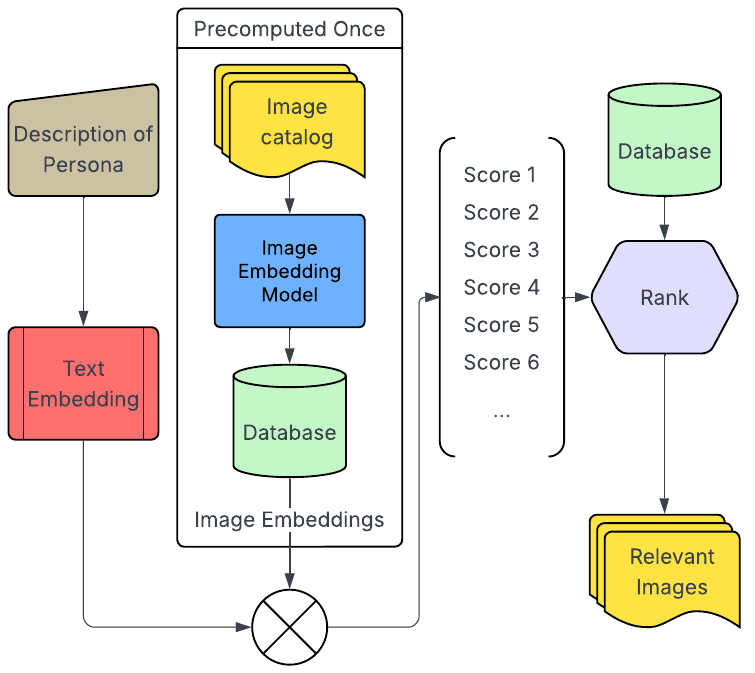}
        \caption{Efficient inference with pre-computed image embeddings.}
        \label{fig:inference}
    \end{figure}

    Our approach (Figure~\ref{fig:inference}) enables inference through a single vector search. Image embeddings are pre-computed once, and at test time only a single text embedding for the query description is required. This yields tractable inference in real-world settings. The overall architecture mirrors standard embedding-based retrieval systems \cite{CLIP, jia2021scalingvisualvisionlanguagerepresentation, zhai2023sigmoidlosslanguageimage}, but our contribution lies in a distillation method that transfers arbitrary preference signals from a teacher vLLM into the embedding space.

    \subsection{Distillation Approach}
    
    Given a persona description $x$ and an image $u$, a relevance score can be  defined as,
    \begin{equation}
        s(x, u) = \mathbf{g}_{\text{text}}(x;\boldsymbol{\theta}_{\text{text}})^{\top} \mathbf{g}_{\text{img}}(u;\boldsymbol{\theta}_{\text{img}}),
    \end{equation}
    where $\mathbf{g}_{\text{text}}$ and $\mathbf{g}_{\text{img}}$ are text and image embedding functions (normalised) with parameters $\boldsymbol{\theta}_{\text{text}}$ and $\boldsymbol{\theta}_{\text{img}}$. We use a pre-trained multi-modal embedding model, freeze $\boldsymbol{\theta}_{\text{text}}$, and fine-tune $\boldsymbol{\theta}_{\text{img}}$ so that $s(x,u)$ reflects the teacher vLLM’s ranking preferences. Training proceeds iteratively. At each step, we sample a batch of $M$ persona descriptions $\mathcal{X}_s = \{x_{s,1}, \dots, x_{s,M}\}$ and, for each persona, a set of $N$ candidate images $\mathcal{U}_s \subset \mathcal{U}$. The teacher vLLM then ranks the $N$ images in $\mathcal{U}_s$, producing an ordering $\{r_{s,1}, \dots, r_{s,N}\}$ by relevance to $x_s$.  
    
    A full ranking is equivalent to $\binom{N}{2}$ pairwise comparisons. We model these comparisons with the Bradley–Terry framework~\cite{bradley-terry}. For two images $u_i$ and $u_j$, the probability that $u_i$ is preferred over $u_j$ for persona $x_s$ is,
    \begin{equation}
            P_{s}(i \succ j) = \frac{\exp(s(x_s,u_i))}{\exp(s(x_s,u_i)) + \exp(s(x_s,u_j))}.
    \end{equation}
    Let $y_{s,ij} = \mathbb{I}\!\left(r_{s,i} < r_{s,j}\right)$ denote the teacher’s binary preference. The pairwise loss is,
\begin{equation}
\begin{split}
\mathcal{L}_{s,ij} = &- y_{s,ij} \log P_s(i \succ j) \\
&- (1 - y_{s,ij}) \log \big(1 - P_s(i \succ j)\big).
\end{split}
\end{equation}

    The total training loss aggregates over all sampled personas and image pairs, $\mathcal{L}(\boldsymbol{\theta}) = \sum_{s=1}^M \sum_{i < j} \mathcal{L}_{s,ij}$, with parameters $\boldsymbol{\theta} = \{\boldsymbol{\theta}_{\text{text}}, \boldsymbol{\theta}_{\text{img}}\}$.
    
    \subsection{Preference-Aligned Distillation} \label{sec:method:preference}

\begin{figure*}[t]
        \centering
        \includegraphics[width=0.7\linewidth]{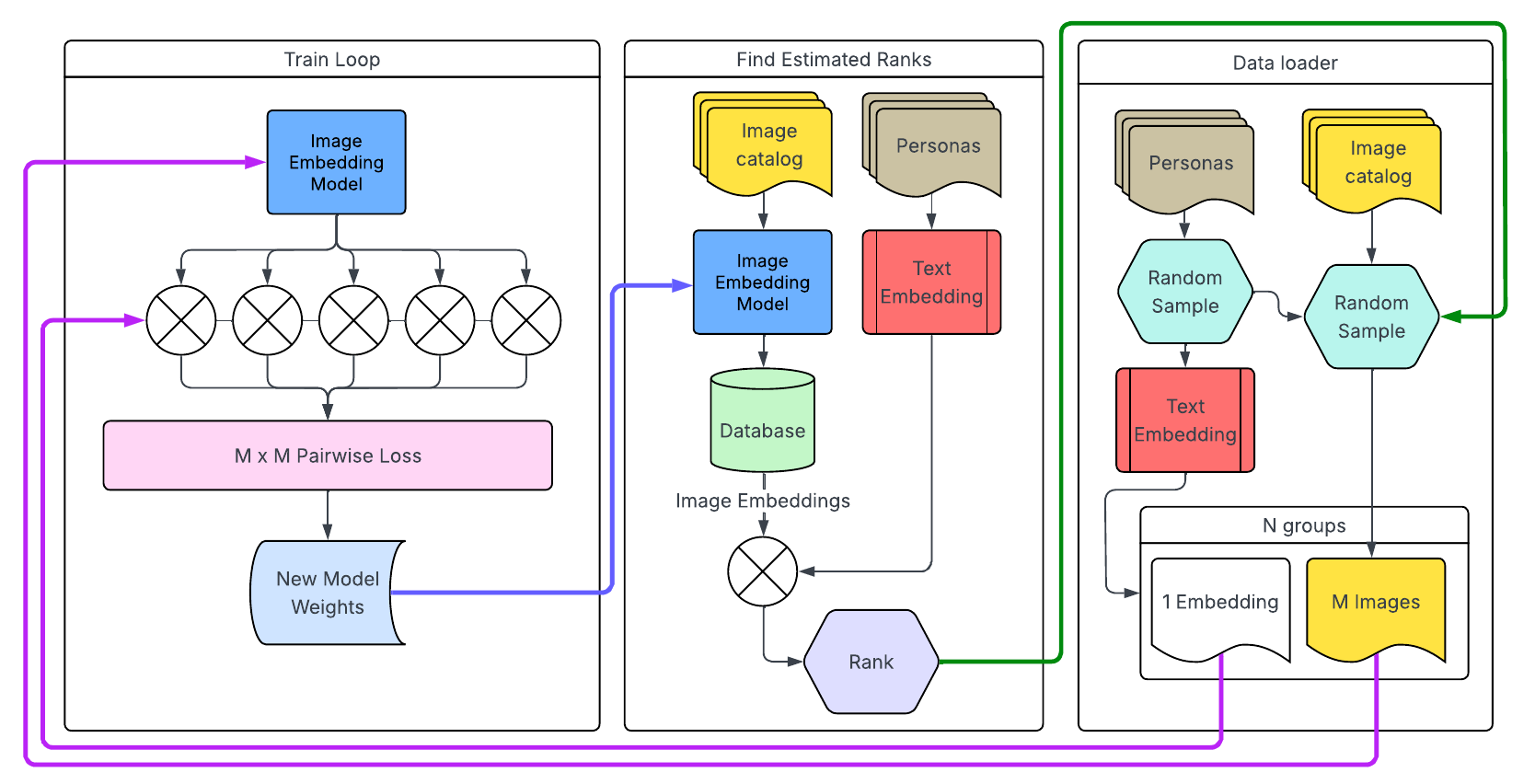}
        \caption{Preference-aligned distillation pipeline}
        \label{fig:training}
    \end{figure*}

    Naive uniform sampling of $\mathcal{U}_s$ is inefficient, since most randomly chosen images are irrelevant to a given persona, yielding uninformative supervision. We therefore introduce a preference-aligned distillation approach, which uses the current student model to guide sampling. At each step, the model computes preliminary relevance scores $s(x,u)$ for all images, which are then partitioned into $P$ bins of varying relevance,
\[
    \mathcal{B}_k(x) = \{ u_i \in \mathcal{U} \mid s(x,u_i) \in I_k \}, k = 1, \dots, P
\]
where $I_1, \dots, I_P$ define disjoint score intervals. Candidate sets $\mathcal{U}_s$ are drawn from a mixture of bins, ensuring diversity between relevant and irrelevant images. This targeted sampling yields more informative vLLM rankings, speeding-up distillation of the teacher’s preferences into the student embedding model. This approach is illustrated in Figure~\ref{fig:training}.
    
    \subsection{Tournament Style Evaluation Labels} \label{sec:tournament-style-label}

    To evaluate our model, we require ground-truth labels indicating the best image for a given persona according to the teacher vLLM. Since obtaining these labels for large catalogs is infeasible, we approximate them with a tournament-based elimination procedure. Given an initial set $\mathcal{U}_{\text{test},0} = \{u_{1,0}, \dots, u_{N,0}\}$ with $N=2^k$, we form disjoint pairs $P_0 = \{(u_{i,0}, u_{i+1,0})\}_{i=0}^{2^{k-1}}$. At each step \(m = 1 \ldots, k\), the teacher vLLM indicates the more relevant image within every pair, halving the candidate pool,
\[
    \mathcal{U}_{\text{test},m} = \{ u \mid (u,v) \in P_{m-1},\; u \succ v \} 
\]
After $N-1$ comparisons, a single top image remains, which we take to be the most preferred image. This provides teacher-derived labels for evaluation on held-out test and validation personas.

\section{Experiments}

\begin{table*}[h]
\centering
\setlength{\tabcolsep}{3pt}
\begin{tabular}{ll|cccc}
\toprule
Persona & Catalog & FashionCLIP (\%) & CLIP (\%) & Text (\%) & Ours (\%) \\
\midrule
Nemotron      & Farfetch  & 83.08 & 81.82 & 74.23 & \textbf{87.31} \\
OpenCharacter & Farfetch  & 81.91 & 78.54 & 81.48 & \textbf{83.93} \\
Nemotron      & H\&M      & 74.45 & 72.35 & 67.32 & \textbf{79.80} \\
OpenCharacter & H\&M      & 83.76 & 79.90 & 73.33 & \textbf{86.10} \\
Nemotron      & UT-Zap50K & 83.79 & 74.19 & 72.40 & \textbf{85.44} \\
OpenCharacter & UT-Zap50K & 80.60 & 69.05 & 64.61 & \textbf{84.21} \\

\bottomrule
\end{tabular}
\caption{Mean percentile rank of the winning image (higher is better) across unseen persona–catalog test sets.}
\label{tab:results}
\end{table*}

    \subsection{Experimental setup}
    
   \paragraph{Datasets.}
We evaluate on two persona datasets: OpenCharacter Personas~\cite{OpenCharacter-Personas} (17,647 train / 50 val / 50 test) and Nvidia's Nemotron Personas~\cite{nvidia_Nemotron-Personas} (5,000 train / 50 val / 50 test). Evaluation is restricted to 50 validation and test personas due to the cost of obtaining tournament-based labels (§\ref{sec:tournament-style-label}). Three product catalogs are used as image datasets: Farfetch Listings~\cite{Farfetch-Listings} (46,910 train / 4,096 val / 4,096 test), the H\&M catalog~\cite{h_and_m} (12,299 train / 4,096 val / 4,096 test), and the UT-Zap50K (shoe images) dataset~\cite{shoe1, shoe2} (6,779 train / 4,096 val / 4,096 test).

    \paragraph{Teacher and Student Models.}
Gemini-2.0-flash~\cite{Gemini} serves as the teacher vLLM, selected for its accuracy and efficiency~\cite{MMMU}. We initialize the student embedding model from FashionCLIP~\cite{fashion-clip}, a CLIP variant trained for fashion understanding, and fine-tune its image encoder using our distillation framework. 
    
\paragraph{Training Details.}
At each step, we sample 1,000 groups of 5 candidate images per persona for teacher labeling. 
Candidate images are chosen using the preference-aligned binning strategy introduced in Section~\ref{sec:method:preference}. 
Concretely, given the lowest and highest student scores $a$ and $b$, we partition the range into four intervals: \(I_1 = [a, \; 0.7a + 0.3b)\), \(I_2 = [0.7a + 0.3b, \; 0.9a + 0.1b) \), \(I_3 = [0.9a + 0.1b, \; 0.95a + 0.05b) \), and \(I_4 = [0.95a + 0.05b, \; b]\). From each group, we sample one image from $I_1$, $I_2$, and $I_3$, and two from $I_4$, ensuring a mix of highly relevant and distractor images. 
Optimization uses AdamW~\cite{AdamW} with an initial learning rate of $1\times 10^{-6}$, decayed by 0.95 per step, and default parameters $\beta_1=0.9$, $\beta_2=0.999$, and $\epsilon=10^{-8}$. 
We use a batch size of 50 with gradient accumulation over 10 steps, and apply early stopping with a patience of 5.

    \paragraph{Baselines.} 
We compare our method against three embedding-based approaches: 
(1) \textbf{FashionCLIP}~\cite{fashion-clip}, a CLIP variant fine-tuned on fashion data, which serves as a strong in-domain baseline; 
(2) \textbf{CLIP (ViT-B/32)}~\cite{CLIP}, the widely used vision--language embedding model trained on large-scale caption–image data; and 
(3) \textbf{Text-only search}, where free-form product descriptions are first generated by Gemini for each image and then embedded with OpenAI’s text-embedding-ada-002 model~\cite{openai2023textembeddingada002}. Retrieval is performed by ranking catalog descriptions according to cosine similarity with persona embeddings. This baseline represents the use of pure text embeddings without leveraging image features, and struggles when catalog descriptions do not fully capture persona-driven attributes.

    \subsection{Experimental results}

    Performance is reported as the mean percentile rank of the teacher-preferred image (Section~\ref{sec:tournament-style-label}), where $100\%$ indicates perfect retrieval. Table~\ref{tab:results} summarizes the results for all dataset combinations.

Our method consistently outperforms all baselines. Against FashionCLIP, the strongest baseline, we obtain improvements of +2\% to +5\% points. The text-only search baseline in particular performs poorly, reflecting both the fundamental mismatch between the embeddings of product and persona descriptions and the inability for text to fully capture products. We also observe differences between the two persona datasets. OpenCharacter personas yield higher overall baseline scores than Nemotron personas, possibly due to clearer textual cues which come with being shorter and more specific. Nemotron personas on the other hand, are broader, and describe many different sides to a persona who might have many unrelated hobbies or traits, making it more more challenging to deduce what images best align with them.

Similarly, retrieval on both Farfetch and UR-Zap50K images is easier than H\&M, likely because both of these contain more varied product styles, creating stronger preference signals for the teacher. UT-Zap50K, which contains only shoes performs roughly as well as Farfetch, likely because despite being limited to a single type of product, the personality types represented contains a lot of variation. In all cases, our approach narrows the gap between scalable embedding-based retrieval and the nuanced reasoning of vLLMs, demonstrating that preference distillation generalizes across both persona sources and product catalogs.

\section{Conclusions and Future Work}

We introduced an efficient framework for distilling the preferences of powerful and flexible vision--language models (vLLMs) into embedding-based retrieval models. By supervising embeddings with teacher-derived rankings, our method enables scalable catalog retrieval while preserving the nuanced, persona-driven alignment of the teacher. Experiments show consistent gains over baselines such as FashionCLIP, CLIP, and text-only search. The approach is not limited to persona-based product recommendation, and can transfer diverse preference signals such as sentiment, style, or user intent. 

Future work includes extending to other domains (e.g., furniture, art, lifestyle), handling multi-turn or conversational queries, and exploring combinations with retrieval-augmented or controllable generation. These directions could yield systems that not only retrieve relevant items but also provide interpretable, personalized recommendations.

\section{Limitations}

\paragraph{Style of personas.} In theory, our methodology could generalize beyond this single style of persona, and we have experimented with another set of personas, the Nvidia Nemotron personas, but have not included them in this study due to time constraints, but will include the result in further work.
\paragraph{Product catalogs.} It is possible for our methodology to apply to other product catalog types beyond fashion.

\paragraph{Compute.} To allow for evaluation, it is required that test and validation sets need to be generated by running tournaments using vLLMs to compare images. This can be slow and costly, but is necessary as a method to fairly measure performance. We also require a large number of inferences from the teacher vLLM to distill it's performance as a one-off cost during training time.

\paragraph{Language.} Our method has been built only for inference using English, but it is also possible to train on other languages or a mix of languages to allow for inference on other languages as well.

\bibliography{custom}

\newpage

\appendix

\section{Risks and Ethics}

There are no perceived risks or ethical concerns associated with this work.

\section{Licensing}
This work is conducted on datasets that are either publicly available or authorized for research use.

The OpenCharacter Personas \cite{OpenCharacter-Personas} are licensed under the Apache License 2.0. The Nvidia Nemotron Personas \cite{nvidia_Nemotron-Personas} as licensed under Creative Commons Attribution 4.0. The H\&M catalog dataset~\cite{h_and_m} is made available on the official H\&M Kaggle page and available for research purposes. The UR-Zap50K dataset is made available on the official University of Texas website for academic purposes. The Farfetch dataset was obtained from Hugging Face \cite{Farfetch-Listings}, where no explicit license is indicated. As per the dataset description, it was third-party scraped and so we do not redistribute, to align with the original dataset terms and conditions. All the models we used in this paper are also authorized for research use. OpenAI's CLIP \cite{CLIP} is licensed under the MIT license. OpenAI’s text-embedding-ada-002 model~\cite{openai2023textembeddingada002} is a proprietary model which we have used according to their terms. FashionClip \cite{fashion-clip} is also licensed under the MIT license. Google Gemini is another proprietary model which we have used according to the terms and conditions \cite{Gemini}. Overall we made our best effort to use the models and datasets as intended by the providers.

\section{Extended Related Work} \label{sec:related-work}

\paragraph{Vision--Language Models.}
Transformer-based vision--language models (vLLMs) such as Flamingo~\cite{alayrac2022flamingovisuallanguagemodel}, LLaVA~\cite{NEURIPS2023_6dcf277e}, Gemini~\cite{Gemini}, and Qwen-VL~\cite{Qwen} have demonstrated impressive performance across a variety of multi-modal reasoning tasks. Recent benchmarks highlight their ability to align text with sets of images but also expose limitations when catalog sizes exceed context window lengths~\cite{MileBench}. Our work targets precisely this setting, where direct use of a vLLM is infeasible due to context constraints.

\paragraph{Embedding-Based Retrieval.}
Contrastive embedding models such as CLIP~\cite{CLIP}, ALIGN~\cite{jia2021scalingvisualvisionlanguagerepresentation}, Florence~\cite{yuan2021florencenewfoundationmodel}, and SigLIP~\cite{zhai2023sigmoidlosslanguageimage} map text and images into a shared space for retrieval. These approaches are efficient at large scale but typically rely on surface-level alignment between text and image content, and thus perform poorly when queries involve abstract attributes (e.g., ``reserved mother'' rather than literal descriptions). Several works have explored fine-tuning embeddings for domain adaptation~\cite{luo2021clip4clipempiricalstudyclip}, though constructing labeled datasets in such settings remains challenging. Our work considers real-world scenarios where there exists no explicit dataset matching persona descriptions to appropriate product images, as may be found in product catalogs.

\paragraph{Distillation of Preferences.}
Knowledge distillation from large models into smaller, efficient students is well studied~\cite{hinton2015distillingknowledgeneuralnetwork}. More recently, preference distillation has been explored in natural language~\cite{rafailov2024directpreferenceoptimizationlanguage} and multi-modal contexts~\cite{NEURIPS2023_6dcf277e}. RankGPT~\cite{luo2024recrankerinstructiontuninglarge} distills pairwise ranking signals from GPT-4 into smaller ranking models, conceptually similar to our approach where vLLM rankings supervise embedding-based models. A component of our contributions is to extend this distillation to vision–language preference ranking with a specific focus on the student being an embedding-based system, which allows for highly efficient ranking at inference time via simple cosine similarity computations.

\paragraph{Personalized and Abstract Recommendation.}
Multi-modal recommendation systems increasingly use both product images and textual descriptions~\cite{Zhang_2019}. While prior work has emphasized literal matching between descriptions and product images, fewer works have studied abstract persona-driven matching. Our problem setup aligns more closely with real-world recommendation scenarios, where users’ descriptions often encode non-literal preferences.

\paragraph{Active Learning and Hard Negative Mining.}
Our iterative data generation process, which queries the vLLM only on subsets of informative images, can be related to active learning~\cite{settles2009active}. In particular, it resembles \textit{active preference learning}, where pairwise queries are selectively sampled to improve a ranking model efficiently~\cite{YUE20121538}. This is also connected to hard-negative mining approaches in metric learning~\cite{Schroff_2015}, where difficult examples accelerate embedding space refinement.

\section{Computational Resources and Cost} \label{sec:compute}

Our approach does require one-off training-time computational resources, primarily due to reliance on Gemini API calls for supervision. During training, 1,000 groups of 5 candidate images per persona are ranked by Gemini, resulting in 1,000 API calls per step. With 8-way parallelism, each step takes about 5 minutes. 
The student embedding model (CLIP, $\sim$150M parameters) is trained on a single RTX 3090 Ti GPU, for which compute demands are modest compared to the API calls.

The largest overhead comes from tournament-style evaluation. Ranking 4,096 images for a single persona requires 8,191 Gemini comparisons. For 50 validation and 50 test personas, this amounts to 819,100 API calls, which takes roughly 12 hours in practice. Thus, while GPU training is efficient, evaluation is dominated by the cost of large-scale Gemini queries.

\begin{figure}[H]
    \centering
    \begin{subfigure}{1\linewidth}
        \includegraphics[width=1\linewidth]{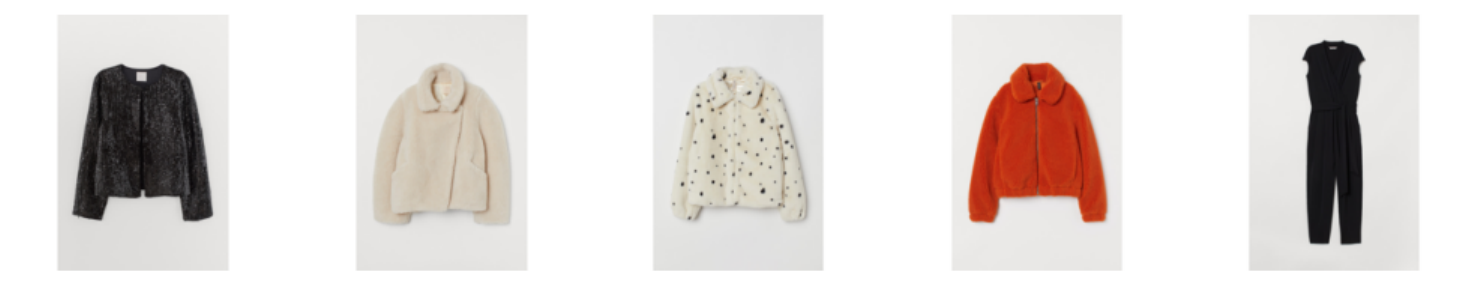}
        \subcaption[]{Top images returned by text embedding search}
    \end{subfigure}
    \vspace{1em}
    \begin{subfigure}{1\linewidth}
        \includegraphics[width=1\linewidth]{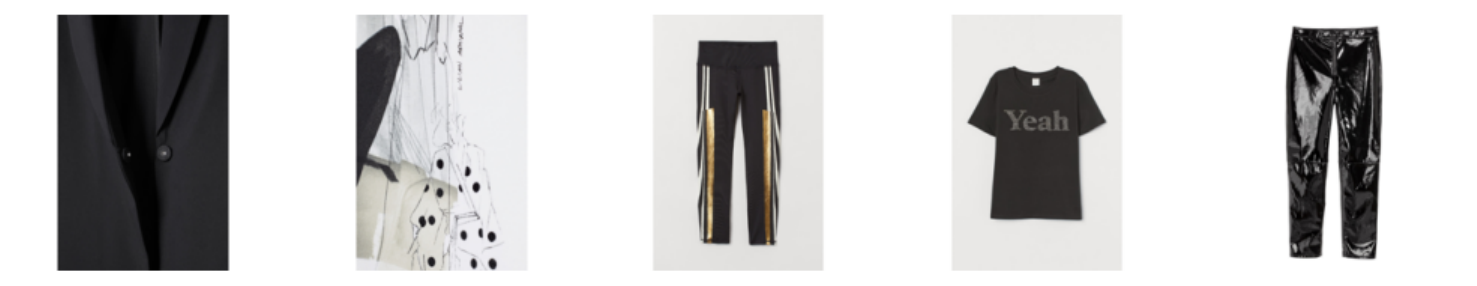}
        \subcaption{Top images returned by CLIP}
    \end{subfigure}
    \vspace{1em}
    \begin{subfigure}{1\linewidth}
        \includegraphics[width=1\linewidth]{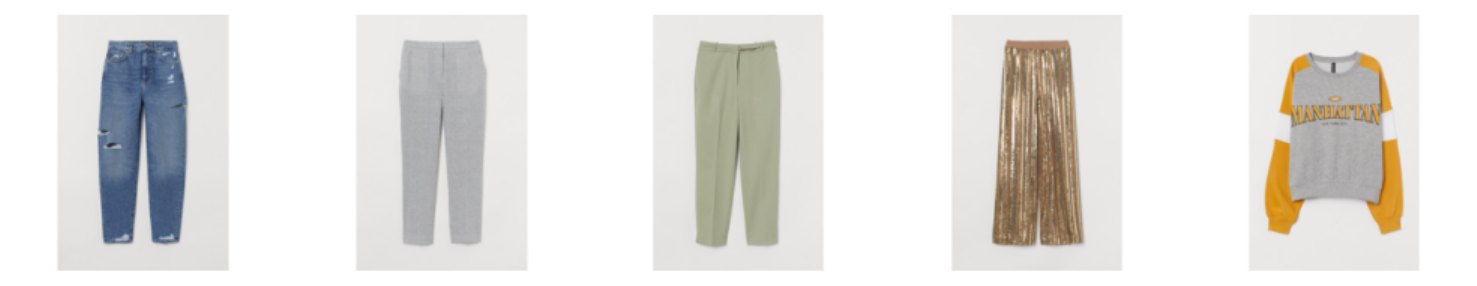}
        \subcaption{Top images returned by FashionCLIP}
    \end{subfigure}
    \vspace{1em}
    \begin{subfigure}{1\linewidth}
        \includegraphics[width=1\linewidth]{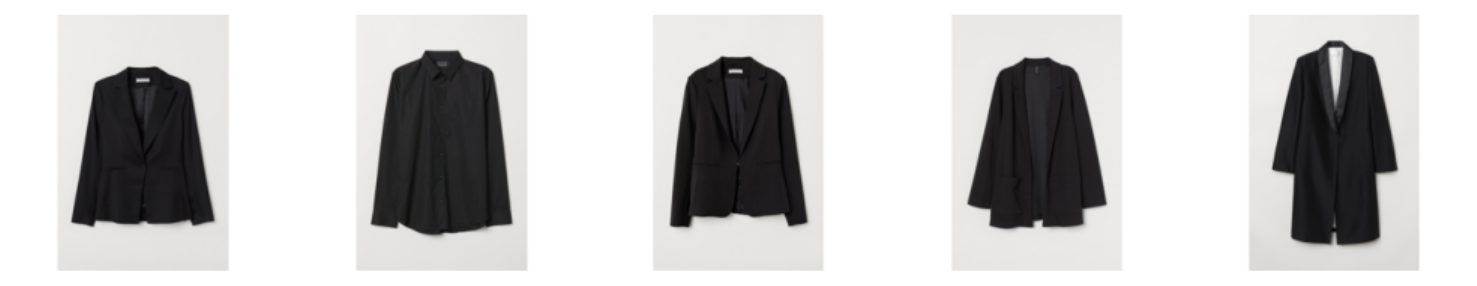}
        \subcaption{Top images returned by \textbf{our} method}
    \end{subfigure}
    \caption{Top 5 images (in descending order of relevance from left to right) out of the test set of 2048 images from the H\&M dataset~\cite{h_and_m} for our method and baseline methods. With the prompt: ``a financial analyst who is skeptical about celebrity wealth estimations" (from OpenCharacter Personas~\cite{OpenCharacter-Personas}).}
    \label{fig:result_images}
\end{figure}

\section{Image Retrieval Example} \label{sec:demonstration}

To demonstrate the differences between different approaches in returning images when given a persona prompt, we include some examples of H\&M catalog~\cite{h_and_m} images recommended by each approach in Figure~\ref{fig:result_images} for an OpenCharacter persona~\cite{OpenCharacter-Personas}. It is interesting to note that our approach returns mostly images of suits and formal-wear given the prompt: ``a financial analyst who is skeptical about celebrity wealth estimations", whilst other approaches fail to do so.

\section{AI Use} \label{sec:ai_use}
AI tools such as copilot were used to generate / auto-complete sections of code, or make minor edits or grammatical checking. All substantive work was written and verified by the authors.

\section{Documentation of Artifacts} \label{sec:documentation}

Google Gemini is a multi-modal large language model developed by Google DeepMind \cite{Gemini}, capable of processing text, images, audio, video, and code. It supports over 100 language and was trained on diverse web and curated datasets, providing broad coverage of domains such as natural language understanding, reasoning, and multi-modal content interpretation. Although it does not encode explicit demographic information, its outputs may reflect biases present in the training data, and performance can vary across underrepresented languages and modalities.

OpenAI's text-embedding-ada-002 \cite{openai2023textembeddingada002} is a proprietary embedding model designed to convert text into high-dimensional vector representations for tasks such as semantic search, clustering, and classification. It was trained on a large and diverse corpus of primarily English-language text covering multiple domains. Its outputs may reflect biases present in the training data, which could be skewed toward Western text, and usage is subject to OpenAI's API terms and policies.

CLIP is an embedding model developed by OpenAI \cite{CLIP}. It maps text and images into a shared embedding space for the purpose of retrieval, classification and zero shot tasks. It is publicly released on GitHub and Hugging Face.

FashionCLIP is a fine tuned version of CLIP developed by Patrick John Chia et al. \cite{fashion-clip} publicly released on Hugging Face. It has been fine-tuned specifically to have a more powerful understanding of fashion and clothing. The model is primarily trained on images labeled in English and could be biased towards Western fashion brands.

The OpenCharacter dataset \cite{OpenCharacter-Personas} consists of synthetic personas designed to be as varied and unbiased as possible. It includes persona descriptions, structured character attributes, and synthetic dialogue between personas, though in this work we only use the persona descriptions. The dataset is publicly available on Hugging Face. While the demographics are randomized, they may still be skewed toward North American or European contexts.

The Nemotron Personas dataset \cite{nvidia_Nemotron-Personas} is a dataset of synthetic personas that are distributed according to real demographic, geographic and personality distributions. The main purpose of the dataset is to capture personas in a broad and unbiased way. It is made available on the official NVIDIA Hugging Face.

The Farfetch listings dataset \cite{Farfetch-Listings} is a publicly available dataset of images originally scraped from the Farfetch website. The images are mostly of products popular in a North American or European context and thus could be biased.

The H\&M catalog dataset \cite{h_and_m} is a publicly available dataset of images available on the official H\&M Kaggle account. It features images of products mostly geared towards a Western audience and thus could also be biased. 

The UT-Zap50K dataset \cite{shoe1, shoe2} is publicly released for academic non-commercial use on the University of Texas website.

\end{document}